# GS-PQM: A Parameter-Domain Quality Metric for Compressed Gaussian Splatting

Pedro Martin, António Rodrigues, João Ascenso, and Maria Paula Queluz

Instituto de Telecomunicações, Instituto Superior Técnico, University of Lisbon, 1049-001 Lisbon, Portugal
{pedro.martin, antonio.rodrigues, joao.ascenso, paula.queluz}@lx.it.pt

***Abstract*—Recent advances in Gaussian Splatting (GS) compression have enabled substantial reductions in GS model size. Reliable objective quality assessment is therefore essential for comparing compression methods and guiding the development of more efficient GS codecs. Existing GS quality assessment typically relies on image and video quality metrics, requiring rendering of predefined viewpoints and making the quality estimate dependent on the selected views. This paper introduces GS-PQM, a novel full-reference quality metric for post-training GS compression that operates directly in the GS parameter domain. GS-PQM estimates perceptual quality from a set of parameter-domain distortion errors using a Support Vector Regression model. Experimental results show that GS-PQM outperforms 25 existing image, video, and point-cloud quality metrics in assessing compressed GS content, providing an accurate and computationally efficient alternative to rendering-based quality assessment.**

## I. Introduction

Gaussian Splatting (GS) [1] has recently emerged as one of the most effective radiance field representations for novel view synthesis and 3D reconstruction, combining high visual quality with real-time rendering. These characteristics make GS particularly attractive for immersive multimedia applications, including virtual and augmented reality, cultural heritage, and telepresence. However, GS models typically contain millions of Gaussian primitives, resulting in large storage requirements that challenge deployment. To address this problem, several GS codecs have recently been proposed based on different coding approaches [2]. While some strategies integrate compression into the training process, post-training codecs are becoming increasingly important in practical scenarios where the original training images, or model creation pipeline, are unavailable.

Current quality assessment of compressed GS models relies almost exclusively on conventional image and video quality metrics [2]. These metrics estimate quality from rendered images or videos synthesized from predefined viewpoints. This approach presents several limitations: (*i*) the estimated quality depends on the selected viewpoints and rendering configuration; (*ii*) rendering multiple images may considerably increase the computational cost of quality assessment; and (*iii*) quality assessment from rendered images does not directly exploit the rich geometric and appearance information encoded in the GS representation itself. Moreover, recent works [3], [4] indicate that existing objective metrics exhibit limited agreement with human perception when applied to compressed GS content.

This paper proposes GS-PQM, a novel full-reference (FR) GS parameter-domain quality metric specifically designed for the quality assessment of post-training GS compression. Instead of operating in the rendered image domain, GS-PQM directly computes distortion from the parameters of the uncompressed and compressed GS models. Thus, the proposed framework introduces GS-Dist, a distortion measure that compares Gaussians, from the compressed and uncompressed models, within a spatially adaptive neighborhood, thereby accommodating coding solutions that may modify, displace, or remove individual Gaussians. GS-Dist considers selected geometric and spherical-harmonic (SH) parameters as features, and the resulting distortion values are mapped to subjective quality scores using a Support Vector Regression (SVR) model. GS-PQM is designed as a perceptually meaningful objective metric for guiding the development, optimization, and evaluation of post-training GS codecs. Experimental results on the GScomp-QA dataset [3] demonstrate that GS-PQM achieves higher correlation with subjective quality scores than the 25 evaluated benchmark metrics, which predominantly assess quality from content rendered from the GS models. The main contributions of this paper can be summarized as follows:

- **GS-PQM metric:** a novel FR objective quality metric[1] for post-training compressed GS models that predicts perceptual quality directly in the GS parameter space, without requiring rendered viewpoints.
- **GS-Dist algorithm:** a novel parameter-domain feature extraction algorithm (used within GS-PQM) that quantifies compression-induced degradations across different GS parameter types.

The remainder of this paper is organized as follows: Section II reviews the related work, Section III presents the proposed GS-PQM metric, Section IV describes the experimental evaluation methodology and discusses the results, and Section V concludes the paper.

## II. Related Work

Quality assessment of GS-generated content is usually performed by rendering predefined viewpoints and evaluating them with conventional 2D quality metrics, such as PSNR, SSIM, and LPIPS [2]. However, recent subjective studies have questioned their suitability for GS content evaluation. GS-QA [4] evaluated several objective metrics and reported limited agreement with subjective scores for GS-generated views. Similarly, GScomp-QA [3] showed that existing quality metrics exhibit relatively weak correlations with human opinion for videos rendered from compressed GS models. These findings highlight the need for GS-oriented quality metrics.

[1] github.com/pedrogcmartin/GS-PQM

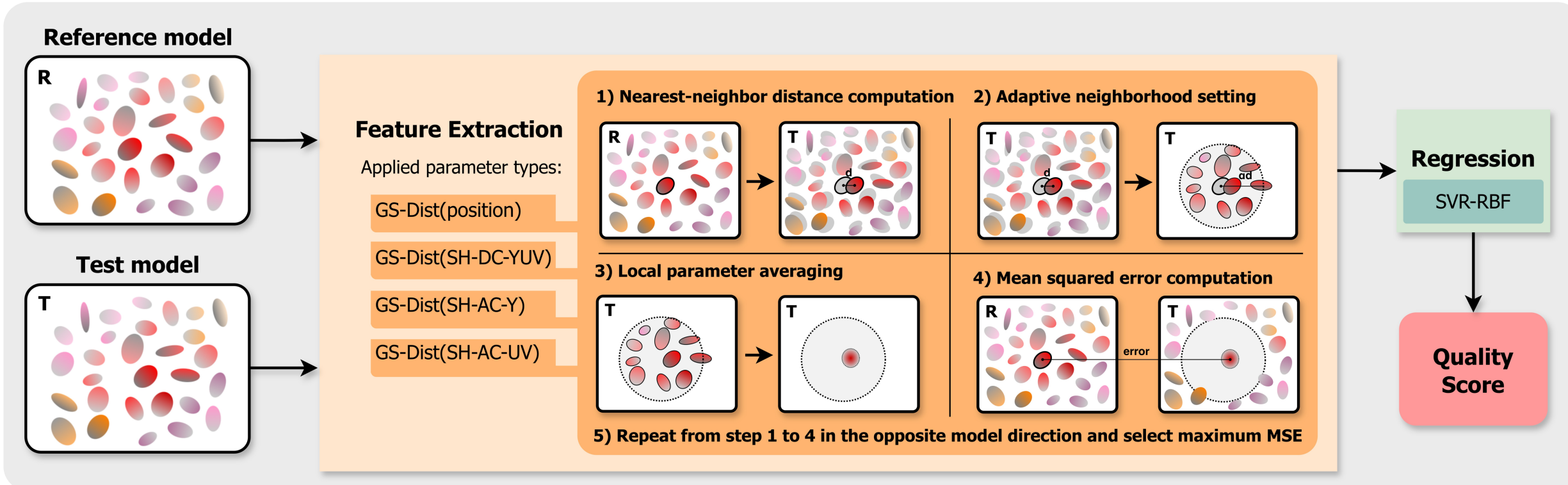

Fig. 1: GS-PQM framework pipeline with an illustration of GS-Dist computation.

To the best of our knowledge, GSOQA [5] is the only objective quality metric specifically developed for GS-generated content. GSOQA is a no-reference (NR) parameter-domain metric that samples local Gaussian neighborhoods, extracts semantic features using a pretrained Gaussian-MAE encoder, models local relationships with Graph Attention Networks, and predicts perceptual quality using an attention-based regression module. Its reported results demonstrate the potential of assessing GS quality directly in the parameter domain. However, GSOQA was evaluated exclusively on synthetic object-centered scenes, and trained using synthetic artifacts (including reduced viewpoints, limited training iterations, Gaussian pruning, and perturbations of Gaussian positions and color) rather than distortions produced by practical GS codecs. Moreover, to the best of our knowledge, its implementation is not publicly available.

In contrast, GS-PQM is a FR parameter-domain quality metric specifically designed, trained, and evaluated using perceptual scores associated with actual compression-induced distortions, on real-world scenes. By directly comparing the parameters of the uncompressed and compressed GS models, it exploits the complementary information provided by both models to estimate perceptual degradation without requiring the rendering of multiple viewpoints.

## III. GS-PQM Framework

This section presents the proposed GS-PQM framework. As illustrated in Fig. 1, GS-PQM first extracts parameter-domain distortion errors by comparing a reference (uncompressed) GS model, denoted by $R$, with a test (compressed) GS model, denoted by $T$, using the proposed GS-Dist error. Fig. 1 also illustrates the feature extraction operations performed on the reference and test models. The resulting features are then combined by a regression model to estimate the perceptual quality of the content rendered from the compressed model.

### A. Feature Extraction

GS-Dist is a parameter-domain distortion measure designed to directly compare the Gaussian primitives of uncompressed and compressed GS models. Rather than establishing a one-to-one correspondence between individual Gaussians from both models (such as MPEG D1 does for point clouds [6]), GS-Dist compares each Gaussian from one model with the average parameter representation of a spatially adaptive neighborhood in the other model. This formulation accounts for compression operations that may modify, displace, or remove individual Gaussian primitives while preserving similar information within their local spatial neighborhood.

Let $N_R$ and $N_T$ denote the number of Gaussian primitives in the reference and test models, respectively. The spatial position of Gaussian $i$ in $R$ is denoted by $\mathbf{x}_i^R$, while $\mathbf{x}_j^T$ denotes the position of Gaussian $j$ in $T$. For a given GS parameter type ($s$), $\mathbf{p}_{i,s}^R$ and $\mathbf{p}_{j,s}^T$ denote the corresponding parameter values associated with these Gaussians. GS-PQM considers four GS parameter types: position, SH DC coefficients, SH AC luminance, and SH AC chrominance. The dimensionality of the parameter vectors depends on the considered parameter type, being three for position and SH DC coefficients, 15 for SH AC luminance, and 30 for SH AC chrominance.

The core concept of GS-Dist is illustrated in Fig. 1 (dark orange) and its computation consists of the following five steps:

1. **Nearest-neighbor distance computation:** For each Gaussian $i$ in $R$, the Euclidean distance to its nearest Gaussian in $T$ is first determined:

$$d_i^{R\to T} = \min_{1\le j\le N_T} \left\|\mathbf{x}_i^R - \mathbf{x}_j^T\right\|_2 \quad (1)$$

where $\|\cdot\|_2$ denotes the L2 norm.

2. **Adaptive neighborhood setting:** A local neighborhood is defined around each reference Gaussian using a radius proportional to its nearest-neighbor distance. Thus, the set of indices of the test Gaussians belonging to the neighborhood of the $i$-th reference Gaussian is defined as:

$$\mathbf{N}_i^{R\to T} = \left\{ j \in [1 \dots N_T] : \left\|\mathbf{x}_i^R - \mathbf{x}_j^T\right\|_2 \le \alpha d_i^{R\to T} \right\} \quad (2)$$

where $\alpha$ is a neighborhood scale parameter controlling the radius and allowing the neighborhood size to adapt to the local Gaussian density.

3. **Local parameter averaging:** For a given GS parameter type $s$, the corresponding values of all test Gaussians within the neighborhood of the reference Gaussian $i$ are averaged as:

$$\bar{\mathbf{p}}_{i,s}^T = \frac{1}{\left|\mathbf{N}_i^{R\to T}\right|} \sum_{j\in \mathbf{N}_i^{R\to T}} \mathbf{p}_{j,s}^T \quad (3)$$

TABLE I. SVR HYPERPARAMETER SEARCH SPACE.

| Hyperparameter | Values |
|---|---|
| Regularization parameter ($C$) | [10, 100, 1000] |
| Kernel coefficient ($\gamma$) | ["scale", 1, 2] |
| Threshold margin ($\varepsilon$) | [0.0001, 0.001, 0.01, 0.1] |

where $|\mathbf{N}_i^{R\to T}|$ denotes the number of test Gaussians contained in the neighborhood of reference Gaussian $i$.

4. **Mean squared error computation:** The distortion for parameter type $s$ is computed as the mean squared difference between the parameter value of each Gaussian in $R$ and the local average of that parameter in the corresponding neighborhood of $T$, according to:

$$\text{MSE}_{R\to T}(s) = \frac{1}{N_R}\sum_{i=1}^{N_R}\left\|\mathbf{p}_{i,s}^{R} - \bar{\mathbf{p}}_{i,s}^{T}\right\|_2^2. \quad (4)$$

5. **Symmetric distortion computation:** Since compression may alter the number and spatial distribution of Gaussian primitives, steps 1 to 4 are repeated in the opposite direction, taking the test model as the source and the reference model as the target. The final GS-Dist error is defined as:

$$\text{GS-Dist}(s) = \max(\text{MSE}_{R\to T}(s), \text{MSE}_{T\to R}(s)) \quad (5)$$

where $\text{MSE}_{T\to R}(s)$ is the mean squared error obtained when repeating steps 1 to 4 from the test to the reference model.

The four resulting GS-Dist errors constitute the input feature vector of the quality regression model (described in Section III.B). GS-Dist errors for other parameter types were also evaluated during feature selection, but consistently yielded lower prediction performance and were therefore not retained. The neighborhood scale was also optimized experimentally, with $\alpha = 5$ providing the best prediction performance.

Moreover, before GS-Dist is computed for the SH coefficients, their RGB components are converted to YUV color space. YUV separates luminance from chrominance information, reflecting the different sensitivity of the human visual system to these components. For the SH DC coefficients, the three YUV components are jointly evaluated in (4), assigning a weight of 1 to the luminance error and 0.25 to each chrominance error to reflect the higher sensitivity of the human visual system to luminance variations. In contrast, for the SH AC coefficients, which encode higher-order directional color information, the luminance and chrominance components are evaluated separately using (4). This allows the regression model to capture their distinct contributions to perceptual quality.

### B. Quality Regression Model

Each GS-Dist feature is independently normalized (using min-max) to ensure comparable numerical ranges before regression. The normalized feature vector is then mapped to a perceptual quality score using an SVR model with a radial basis function (RBF) kernel. The RBF kernel enables the regression model to capture the relationship between parameter-domain distortions and subjective quality. Once trained, the regression model receives the selected and normalized GS-Dist features as input and outputs a single score estimating the perceptual quality of the content rendered from the compressed GS model.

TABLE II. PLCC AND SROCC PERFORMANCE EVALUATION.

| Metric | PLCC | SROCC | Metric | PLCC | SROCC |
|---|---|---|---|---|---|
| GS-PQM | **0.873** | **0.786** | FVVDP | 0.631 | 0.615 |
| MANIQA | 0.756 | 0.762 | LPIPS | 0.623 | 0.588 |
| NLPD | 0.768 | 0.712 | MAD | 0.582 | 0.611 |
| GMSD | 0.747 | 0.727 | DISTS | 0.600 | 0.592 |
| ST-LPIPS | 0.728 | 0.707 | VSI | 0.569 | 0.529 |
| DOVER | 0.734 | 0.681 | CVVDP | 0.558 | 0.538 |
| HYPERIQA | 0.699 | 0.716 | SSIM | 0.555 | 0.528 |
| PSNR-HVS | 0.721 | 0.673 | PSNR-Y | 0.539 | 0.526 |
| FSIM | 0.686 | 0.690 | C3DVQA | 0.486 | 0.577 |
| IW-SSIM | 0.710 | 0.663 | MS-SSIM | 0.520 | 0.529 |
| VIF | 0.710 | 0.659 | MSE-RGB | 0.498 | 0.543 |
| VMAF | 0.649 | 0.632 | PSNR-YUV | 0.459 | 0.460 |
| SimpleVQA | 0.649 | 0.619 | MPEG D1 | 0.079 | 0.204 |

## IV. PERFORMANCE EVALUATION

This section describes the SVR training procedure, the test conditions for the performance evaluation of GS-PQM, and the respective results.

### A. Training Procedure

GS-PQM was trained using the GScomp-QA subjective dataset [3] which, to the best of our knowledge, is the only publicly available dataset providing subjective quality scores for content rendered from compressed GS models. The dataset considers 13 real-world scenes whose uncompressed GS models were generated using the original GS method [1] and subsequently compressed using two post-training codecs, GSICO [7] and FCGS [8], at different compression levels, resulting in 48 compressed models. For each compressed model, the dataset provides a rendered video and its associated Differential Mean Opinion Score (DMOS). For GS-PQM training, each sample is defined by a GS-Dist feature vector, computed between the corresponding uncompressed and compressed models, used as the SVR input, together with the DMOS of the video rendered from the compressed model, used as the regression target.

To evaluate the generalization capability of GS-PQM, a repeated scene-independent four-fold cross-validation protocol was adopted. The 13 scenes were randomly partitioned into four folds while ensuring that all data associated with the same scene, regardless of codec or compression level, were assigned to the same fold. Each fold comprised the data associated with 12 compressed GS models, corresponding to three or four distinct scenes. Consequently, the regression model was always evaluated on scenes that were completely unseen during training, preventing scene leakage and avoiding overly optimistic performance estimates arising from content overlap between the training and test sets. During each cross-validation run, the SVR model was trained using three folds and evaluated on the remaining fold. This process was repeated four times so that each fold served once as the test set.

### B. Test Conditions

The prediction performance of GS-PQM was quantified using the Pearson Linear Correlation Coefficient (PLCC) and the Spearman Rank-Order Correlation Coefficient (SROCC), which respectively measure linear correlation and rank-order monotonicity with respect to the subjective DMOS values. To reduce the influence of a particular random scene partition, the entire cross-validation procedure was repeated 200 times using different random fold assignments. For each repetition, the predictions from the four test folds were first aggregated to compute a single global PLCC and SROCC. The final reported performance corresponds to the average of these global correlation values over the 200 repetitions. Within each cross-validation repetition, the SVR hyperparameters were optimized exclusively using the training data through a grid search over the parameter space summarized in Table I. The optimal configuration was selected independently for each training split before evaluating the corresponding test fold, ensuring that the test data were never used during hyperparameter tuning.

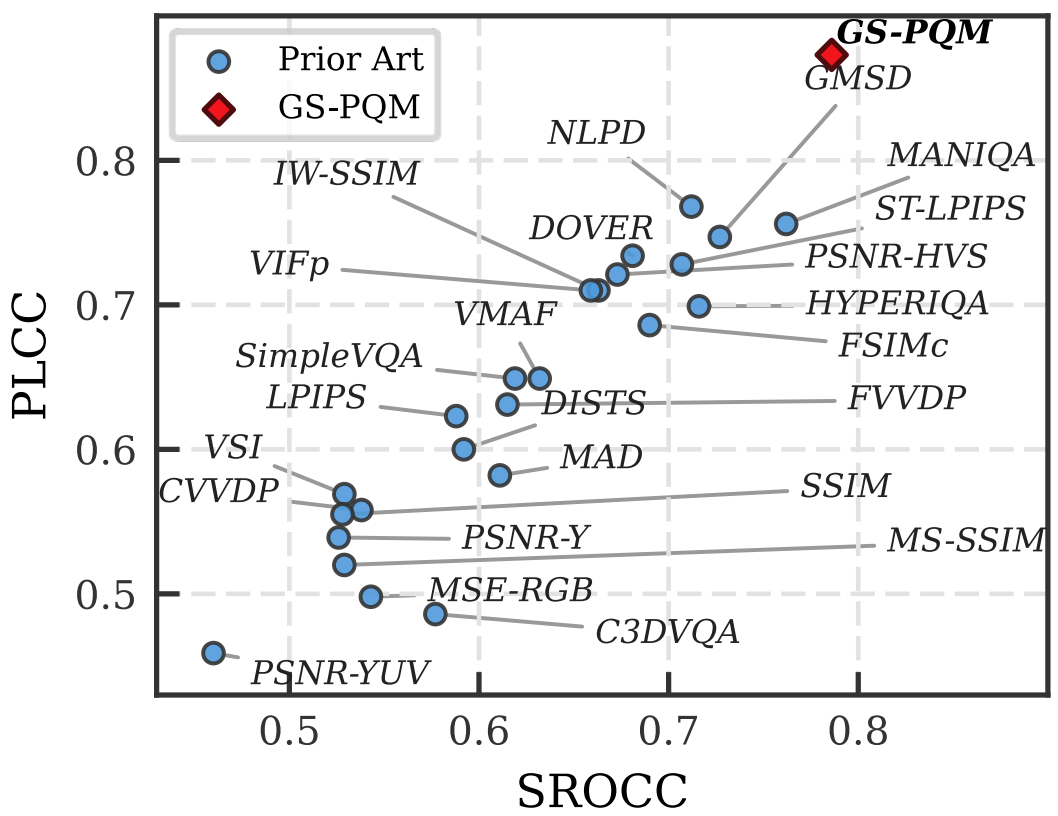

Fig. 2: Correlation performance of GS-PQM against existing objective metrics that work on the rendered domain.

### *C. Experimental Results*

Table II compares the proposed GS-PQM with 25 benchmark objective quality metrics, covering classical fidelity metrics, perceptual image and video quality metrics, recent learning-based methods, and MPEG D1 (a point-to-point metric developed by MPEG for point clouds [6]). For the rendered-domain benchmark metrics, the reference and test images/videos provided by GScomp-QA [3] were used following the evaluation procedure defined in that work. The evaluated methods comprise all metrics considered in GScomp-QA, together with C3DVQA [9], CVVDP [10], DOVER [11], HYPERIQA [12], MANIQA [13], MPEG D1, and SimpleVQA [14]. Fig. 2 provides a visual comparison of the performance results reported in Table II.

GS-PQM achieves the highest correlation with subjective quality among the 26 evaluated metrics, with a mean PLCC of $0.873 \pm 0.022$ and a mean SROCC of $0.786 \pm 0.022$ across the 200 cross-validation repetitions. The closest benchmark in terms of PLCC is NLPD, with 0.768, while MANIQA provides the closest SROCC, with 0.762. These results indicate that the proposed parameter-domain representation provides stronger agreement with subjective quality than the considered rendering-domain and point-cloud-based alternatives. This improvement is particularly noteworthy considering that GS-PQM relies on only four parameter-domain features derived from the proposed GS-Dist error formulation. Rather than extracting complex semantic representations or requiring rendering, the proposed framework directly exploits information available in the compressed and uncompressed GS models.

## V. Conclusion

This paper presented GS-PQM, a novel FR objective quality metric for post-training GS compression that operates directly in the GS parameter domain. The proposed framework introduces GS-Dist to compute parameter-domain distortion errors, which are subsequently mapped to a quality score using an SVR model. Experimental results on the GScomp-QA dataset showed that GS-PQM achieves the highest correlation with subjective scores among 26 evaluated metrics, with a PLCC of 0.873 and a SROCC of 0.786. These results demonstrate the potential of parameter-domain quality assessment for compressed GS models. Furthermore, they support the use of GS-PQM as a perceptually meaningful objective criterion for guiding the development of post-training GS codecs and comparing their quality performance without requiring rendered viewpoints.